# MechAInistic: An LLM-guided Multi-Agent System for Reasoning over Genome-Scale Constraint-Based Metabolic Models

Josh Loecker[1,3,*], Narayna Puraja[1,2], William Bryant[1,2], Bhanwar Lal Puniya[1], Prakash Packrisamy[1], Ahmed Abdeen Hamed[1,4,**], and Tomáš Helikar[1,4,**]

[1]Biochemistry Department, University of Nebraska-Lincoln, Lincoln, NE, US

[2]These authors contributed equally

[3]Lead author: Josh Loecker

[4]Senior authors: Ahmed Abdeen Hamed and Tomas Helikar

*Correspondence: jloecker3@nebraska.edu
**Correspondence: ahamed4@nebraska.edu
**Correspondence: thelikar2@nebraska.edu

## SUMMARY

Constraint-based metabolic modeling is a powerful way to study the mechanistic basis of cellular states and disease, but its effective use demands substantial computational expertise and careful coordination of multi-step analyses. We developed MechAInistic to lower this barrier and enable researchers to ask complex biological questions in natural language. Harnessing large language models, MechAInistic is a multi-agent system organized around an Architect-Reviewer pattern that transforms a natural-language question into an executable, model-grounded workflow and generates a structured report. The system supports a variety of tasks, including pathway comparison, perturbation analysis, drug-target exploration, and literature-grounded interpretation across paired metabolic model states. We developed and evaluated MechAInistic using two paired immune-cell metabolic-model use cases for therapeutic hypothesis generation. For Naive B cells from rheumatoid arthritis (RA) paired with healthy controls, MechAInistic identified mitochondrial metabolic rewiring and nominated Devimistat/CPI-613 as an investigational OGDH-centered hypothesis. In a paired CD4+ Th17 cell study from multiple sclerosis (MS) and healthy controls, the same workflow identified NADP-dependent isocitrate dehydrogenase as the optimal single target and proposed *ivosidenib* as an FDA-approved repurposing candidate. Together, these results show that MechAInistic converts natural-language biological questions into executable, model-grounded workflows for traceable therapeutic hypothesis generation. MechAInistic is accessible at [“https://mechainistic.dtih.org”].

## GRAPHICAL ABSTRACT

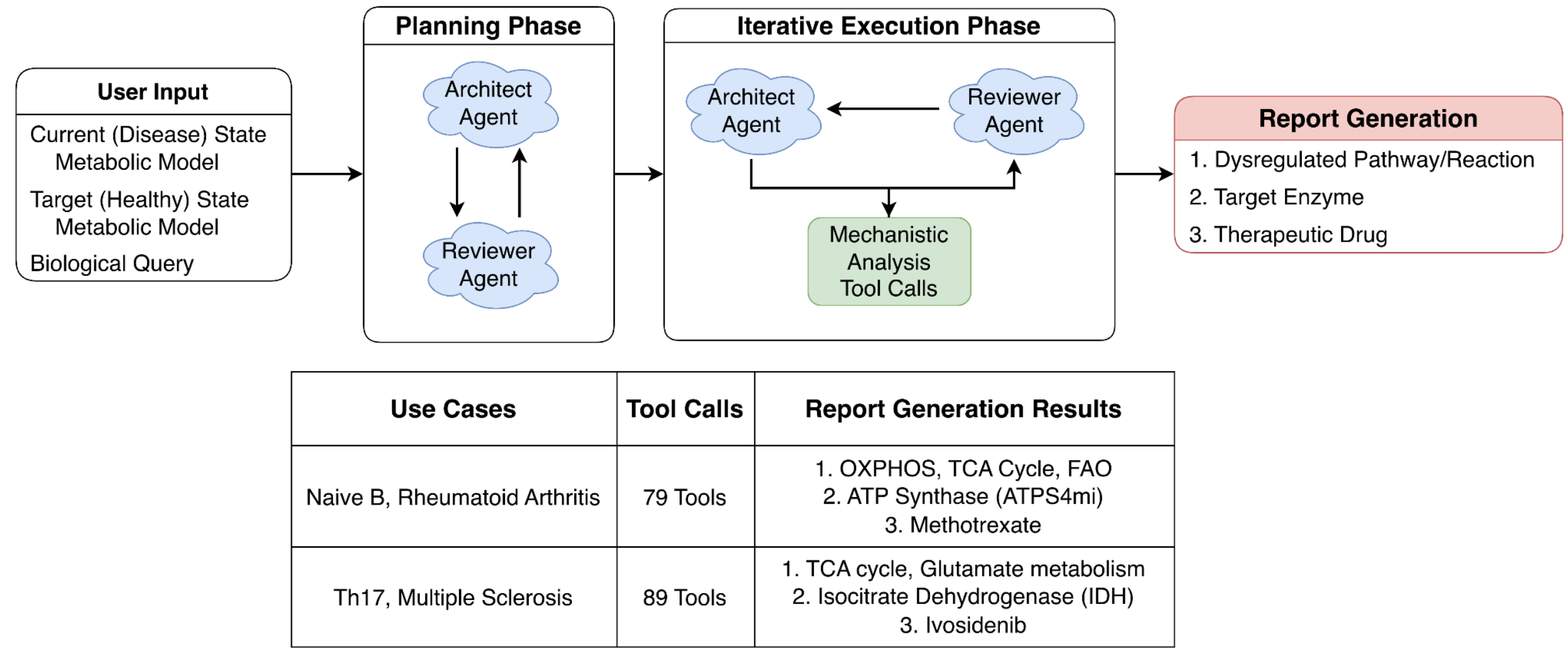

| Use Cases | Tool Calls | Report Generation Results |
|---|---|---|
| Naive B, Rheumatoid Arthritis | 79 Tools | 1. OXPHOS, TCA Cycle, FAO<br>2. ATP Synthase (ATPS4mi)<br>3. Methotrexate |
| Th17, Multiple Sclerosis | 89 Tools | 1. TCA cycle, Glutamate metabolism<br>2. Isocitrate Dehydrogenase (IDH)<br>3. Ivosidenib |

## INTRODUCTION

Constraint-based models (CBMs) have become foundational tools for investigating metabolic capabilities, predicting phenotypic outcomes, and characterizing disease-associated metabolic alterations across biological systems.[1,2] Despite their conceptual accessibility, effective use of CBMs requires computational and biological expertise.[3,4] Frameworks such as COBRA Toolbox, COBRAxy, and COBRApy provide powerful analytical capabilities, but they require proficiency in programming, numerical optimization, model configuration, and metabolic-network interpretation.[5–7] Even routine CBM analyses often require multi-stage workflows that integrate model setup, simulation, perturbation analysis, and interpretation.[1,8,9] As a result, users must assemble analysis pipelines manually, manage tool configurations, and interpret intermediate outputs, creating barriers to reproducibility and limiting accessibility for researchers without specialized computational training.[10–12]

Large language models (LLMs) offer a potential route to more accessible model interaction by enabling natural-language-driven workflows. Recent systems use prompt engineering, in-context learning, retrieval-augmented generation, and agentic AI approaches to assist biological modeling and scientific workflow construction.[13–15] However, LLM-generated workflows can fail under realistic conditions because they are not consistently grounded in executable tools, model constraints, or verifiable intermediate results. For CBM applications, a useful LLM-enabled system must translate biological questions into executable analyses, check intermediate outputs, adapt when tool calls fail or produce incomplete evidence, and preserve traceability from the original question to model-derived results.

To address this need, we developed MechAInistic, an agentic AI system that orchestrates CBM analyses through a reviewer-supervised, tool-grounded workflow. MechAInistic separates workflow design from workflow verification, interfaces directly with CBM tools, evaluates intermediate outputs, and transforms natural-language questions into structured reports containing computational results, mechanistic interpretation, literature support, and limitations. By coupling LLM-based planning and review to executable metabolic-modeling tools, MechAInistic lowers the barrier to CBM analysis while preserving the transparency required for model-grounded biological reasoning.

## RESULTS

We developed MechAInistic as a multi-agent, LLM-guided system for natural-language reasoning over paired CBMs. The Results first present the reviewer-supervised architecture, then demonstrate model-grounded workflow generation, comparator performance against general-purpose LLMs, and two immune-cell therapeutic hypothesis-generation use cases.

### *MechAInistic operationalizes a reviewer-supervised agentic architecture for metabolic-model reasoning*

We developed MechAInistic to reduce the technical barrier between biological questions and executable constraint-based metabolic modeling. The system accepts two paired metabolic models, representing the current and target states, along with a natural-language query. In the use cases presented here, the current state corresponds to a disease-state immune-cell model, and the target state corresponds to the matched healthy reference model. MechAInistic then translates the user's query into a structured computational workflow, executes model-grounded analyses, retrieves relevant literature, and produces a final report that combines quantitative results, mechanistic interpretation, and stated limitations.

The platform is organized around three role-specialized agents. The Architect agent decomposes the user's question, develops an analytical plan, and proposes tool calls against the uploaded metabolic models. The Reviewer agent evaluates the plan and the accumulated evidence before the system proceeds to final synthesis. The Task agent supports PubMed query generation and summarizes tool outputs, enabling quantitative results to be reused during iterative reasoning without overwhelming the context window. This separation of roles was designed to prevent the system from moving directly from language-model inference to biological conclusion without intermediate checks.

MechAInistic grounds its reasoning in executable metabolic-modeling tools rather than language-model text alone. These tools support model simulation, flux-distance calculation, pathway and reaction comparison, graph-based network analysis, perturbation simulation, dose-response analysis, drug-target lookup, robustness analysis, and literature retrieval. The resulting workflow creates an auditable chain from the original biological query to executed tool calls, model-derived quantitative outputs, literature-grounded interpretation, and final synthesis. Detailed implementation information, including LLM configuration, tool parameters, prompt templates, workflow-control logic, and software dependencies, is provided in the Supplementary Information.

### *MechAInistic generates model-grounded workflows from natural-language biological questions*

To demonstrate model-grounded workflow generation, we applied MechAInistic to two paired-model therapeutic hypothesis-generation tasks using the same prompt: "Identify a drug therapy that could restore metabolic flux in the disease-state model to the healthy state. Propose a single drug therapy that maximizes therapeutic efficacy while minimizing off-target metabolic disruption." The first task used rheumatoid arthritis Naive B-cell models, and the second used multiple sclerosis CD4+ Th17-cell models. Disease and cell-type labels were not included in the prompt and were available only through the uploaded model context. Illustrative examples of broader query types that MechAInistic can translate into CBM workflows are provided in Table S1.

In the rheumatoid arthritis Naive B-cell case, MechAInistic loaded the supplied JSON models, calculated baseline metabolic divergence, identified reaction- and pathway-level differences, simulated candidate perturbations, and generated a drug-therapy hypothesis. The disease-state model differed substantially from the healthy reference, with a Euclidean flux distance of 151.60 and cosine similarity of 0.093. The analysis identified increased mitochondrial activity, including elevated oxidative phosphorylation, citric acid cycle activity, and reactive oxygen species detoxification. At the reaction level, MechAInistic prioritized 2-oxoglutarate dehydrogenase/OGDH, represented in the model by AKGDm, because it carried disease-state forward flux of approximately 5 mmol $gDW^{-1}$ $h^{-1}$ with no corresponding flux in the healthy reference model.

In the multiple sclerosis CD4+ Th17 case, MechAInistic again generated model-derived outputs from the paired disease and healthy metabolic models. Baseline characterization yielded a Euclidean flux distance of 25.247 between disease and healthy states. The system identified NADP-dependent isocitrate dehydrogenase, represented by ICDHy and mapped to the IDH1/IDH2 enzyme family, as a disease-associated reaction with forward flux of 3.441 in the disease model and no corresponding flux in the healthy model. Simulated inhibition reduced the Euclidean distance from 25.247 to 23.069 and produced no detectable reduction in the biomass objective across the tested inhibition range. The analysis also identified a broader glutamate-associated phenotype, including a 2.17-fold elevation in glutamate metabolism and reduced citric acid cycle flux relative to the healthy state.

Together, these demonstrations show that MechAInistic can convert natural-language biological questions into executable, model-grounded workflows across distinct disease states, immune-cell types, and metabolic-model contexts. The outputs included model-derived flux distances, reaction identifiers, pathway-level changes, simulated perturbation effects, drug-target hypotheses, and explicit caveats about confidence and validation.

### *MechAInistic demonstrates improved model grounding and task completion over general-purpose LLM systems*

We compared MechAInistic with general-purpose LLM systems to determine whether a dedicated, tool-grounded agentic architecture provides practical advantages over standard language-model workflows for metabolic-model reasoning. Anthropic's Claude, OpenAI's ChatGPT, Microsoft's Copilot, and Google's Gemini were given the same paired metabolic model files and the same prompt used for MechAInistic: "Identify a drug therapy that could restore metabolic flux in the disease-state model to the healthy state. Propose a single drug therapy that maximizes therapeutic efficacy while minimizing off-target metabolic disruption."

We evaluated each system using a nine-axis rubric designed to distinguish plausible biological narrative from model-grounded computational reasoning. The rubric assessed: (1) model fidelity, defined as evidence that the supplied JSON metabolic models were loaded, parsed, and used; (2) methodological reproducibility, including named tools, solvers, scoring rules, or executable analysis steps; (3) literature evidence, including use of traceable peer-reviewed sources; (4) computational evidence, including flux values, perturbation outputs, or distance metrics; (5) direct relevance to the single-drug flux-restoration task; (6) whether the question was answered with a specific single therapy; (7) evidence of structured review or agentic control; (8) reproducibility and execution feasibility, including whether the workflow was rerunnable or limited by file-ingestion, context, rate-limit, or truncation failures; and (9) human-review traceability, including explicit indication of manual review, sign-off, or correction. Detailed scoring criteria are provided in the Methods and Supplementary Information.

Across both use cases, MechAInistic showed the most consistent workflow-level performance. It loaded and operated on the supplied JSON models, reported model-derived reaction fluxes and distance metrics, identified reaction-level control points, simulated perturbations, retrieved supporting literature, and documented the analytical path leading to each recommendation. General-purpose LLM systems produced plausible biological recommendations and, in some cases, sophisticated model-aware reasoning, but their outputs varied substantially across file ingestion, quantitative model-derived evidence, methodological traceability, and task completion.

The comparator analyses also showed that apparent sophistication of a final report was not sufficient evidence of model-grounded correctness. In one Claude run for the multiple sclerosis CD4+ Th17 use case, the final report presented an extensive quantitative analysis and recommended dimethyl fumarate. However, audit of the underlying steps revealed execution-level issues not visible from the final report alone. Claude manually inhibited reactions associated with selected genes rather than applying COBRApy's gene-protein-reaction-aware knockout logic, leading to reaction inhibitions that would not occur when Boolean isozyme relationships were respected. It also mislabeled exchange-flux directionality by not accounting for uptake sign conventions, reported an analysis step that had been skipped, and made an unsupported mechanistic claim linking dimethyl fumarate to monocarboxylate transporter restoration. In addition, Claude used pFBA to evaluate post-perturbation states. Although pFBA is useful for obtaining parsimonious optimal flux distributions, it answers a different optimization question than MOMA/linear MOMA-style perturbation analysis, which explicitly evaluates perturbed solutions relative to a reference flux distribution.

| System | RA Naive B-cell recommendation | MS CD4+ Th17 recommendation |
|---|---|---|
| MechAInistic (Kimi K2.6, Qwen3.5, Gemma-4) | Devimistat / CPI-613 (α-KGDH inhibitor[16]) | Ivosidenib (IDH1 Inhibitor[17]) |
| Claude (Sonnet 4.6 Adaptive) | No recommendation (Error, Out of Credit) | Dimethyl fumarate; extensive response, but audit identified execution-level errors |
| ChatGPT (GPT-5.5 Thinking) | Metformin (ETC Complex I Inhibitor[18]) | Calcitriol (Vitamin D3 Inhibitor[19]) |
| Copilot ("Think Deeper") | Metformin (ETC Complex 1 Inhibitor[18]) | Metformin (ETC Complex 1 Inhibitor[18]) |
| Gemini 3 Flash | No recommendation (Error in responding) | No recommendation (Error in responding) |

**Table 3.** Therapeutic recommendations issued by each system for the two use cases.

Therapeutic recommendations differed substantially across systems: MechAInistic nominated Devimistat/CPI-613 and ivosidenib, whereas general-purpose LLM systems returned metformin, calcitriol, dimethyl fumarate, or no recommendation depending on the interface and use case. We did not treat these differences as evidence that one drug was correct, because the correct therapeutic answer cannot be established computationally. Instead, the divergence illustrates why drug identity alone is an insufficient endpoint. The primary distinction was whether the recommendation was supported by explicit model ingestion, flux-derived quantitative evidence, perturbation simulation, literature grounding, stated limitations, and an auditable path from user query to model-derived interpretation.

### *MechAInistic generates a mitochondrial-metabolism hypothesis in rheumatoid arthritis Naive B cells*

In the rheumatoid arthritis Naive B-cell use case, MechAInistic identified mitochondrial metabolism as a major contributor to the disease-state flux profile. B cells are relevant to rheumatoid arthritis pathogenesis and therapeutic targeting.[20] The disease model showed increased oxidative phosphorylation, citric acid cycle activity, and reactive oxygen species detoxification relative to the healthy reference, suggesting broader mitochondrial rewiring rather than a single isolated reaction change.

MechAInistic evaluated candidate mitochondrial reactions by comparing disease-state and healthy-state fluxes, simulating perturbations, and assessing whether each intervention moved the disease model toward the healthy flux distribution. Initial perturbation tests did not support all candidate mitochondrial targets equally: simulated succinate dehydrogenase inhibition increased the Euclidean distance from 151.60 to 152.44, while fumarate-associated inhibition produced only a small improvement, reducing the distance from 151.60 to 151.44. The system subsequently prioritized 2-oxoglutarate dehydrogenase/OGDH, represented in the model by AKGDm, because it carried disease-state flux but no corresponding healthy-state flux.

MechAInistic proposed that OGDH inhibition could reduce excess citric acid cycle flux and indirectly suppress downstream oxidative phosphorylation and reactive oxygen species detoxification. It nominated Devimistat/CPI-613 because published evidence supports inhibition of OGDH and pyruvate dehydrogenase by CPI-613.[16,21–25] The system further reasoned that concurrent pyruvate dehydrogenase inhibition may be mechanistically compatible with the disease-to-healthy transition predicted by the model, because the healthy reference state showed reduced mitochondrial TCA/OXPHOS activity. Simulated inhibition preserved the biomass objective at 1.6889 despite a moderate negative correlation between OGDH perturbation and biomass maintenance, suggesting network-level compensation in the model.

Manual validation supported the model-derived quantitative outputs and the literature-supported connection between Devimistat/CPI-613 and OGDH/PDH inhibition. However, validation also showed that Devimistat/CPI-613 is investigational and not FDA-approved for any indication. The literature-retrieval and summarization process also produced at least one title-author mismatch despite retrieving biologically relevant PubMed-indexed sources. Thus, this use case demonstrates MechAInistic's ability to generate a model-grounded mitochondrial therapeutic hypothesis while also showing why human review remains necessary for clinical interpretation and citation fidelity.

### *MechAInistic generates an IDH-centered multiple sclerosis Th17-cell hypothesis*

In the multiple sclerosis CD4+ Th17 use case, MechAInistic identified NADP-dependent isocitrate dehydrogenase, represented by ICDHy and mapped to the IDH1/IDH2 enzyme family, as the strongest single-target intervention. CD4+ Th17 cells are central to multiple sclerosis pathogenesis and autoimmune inflammation.[26,27] The disease-state model carried forward ICDHy flux of 3.441, whereas the healthy-state model carried no flux through this reaction. Simulated inhibition reduced the Euclidean distance between disease and healthy flux states from 25.247 to 23.069, corresponding to an 8.6% improvement. ICDHy was not classified as a topological hub, and robustness and dose-response analyses showed no detectable change in the biomass objective across 0–100% inhibition, suggesting a potentially favorable therapeutic window within the model.

The biological interpretation generated by MechAInistic centered on a glutamate-overflow phenotype. In this interpretation, disease-associated ICDHy activity generates α-ketoglutarate that is diverted toward glutamate metabolism rather than continuing through oxidative TCA-cycle activity. This interpretation was consistent with the model-derived increase in glutamate metabolism and reduced citric acid cycle flux relative to the healthy reference. Although the prompt requested a single drug therapy, MechAInistic also explored whether a multi-target perturbation could better restore the healthy flux state. Combined inhibition of ICDHy, r2420, and MALtm improved Euclidean distance to 21.591 and yielded a cosine similarity of 0.984, outperforming ICDHy inhibition alone. Because no single available agent targeted all three nodes, the system retained this result as mechanistic context rather than the final recommendation.

MechAInistic initially stated incorrectly that ICDHy lacked a gene identifier in the model, but its literature-retrieval phase connected ICDHy to the IDH1/IDH2 enzyme family and identified clinically available IDH inhibitors, including ivosidenib and vorasidenib.[28–30] It selected ivosidenib as the primary candidate. Manual review supported the broader IDH-centered mechanism but suggested that vorasidenib may be mechanistically preferable because it inhibits both IDH1 and IDH2, whereas ivosidenib primarily targets IDH1. Literature linking IDH activity, α-ketoglutarate, 2-hydroxyglutarate, and Th17/Treg balance supported the broader target-class hypothesis.[31] However, human review suggested that vorasidenib may be mechanistically preferable to ivosidenib in this context because it inhibits both IDH1 and IDH2, whereas ivosidenib primarily targets IDH1.[32]

This use case illustrates both the strength and limitation of MechAInistic: the system identified a biologically meaningful target class, explored a broader perturbation space, and returned to the monotherapy constraint, but expert review remained necessary for annotation correction and compound-level prioritization. The resulting hypothesis was consistent with literature on Th17 metabolism and TCA-cycle breakpoints.[33,34]

## METHODS AND SYSTEM DESIGN

This section describes the implementation of MechAInistic and the procedures used to generate and evaluate the results reported above. Detailed materials not required for interpreting the main Results, including full prompt templates, tool-parameter definitions, workflow-state variables, software dependencies, accession tables, and comparator outputs, are provided in the Supplementary Information.

### *System implementation and user input*

MechAInistic is implemented as a web-accessible system for executing natural-language queries over paired constraint-based metabolic models. Users provide two COBRA-compatible metabolic models encoded in JSON format: a current-state model and a target-state model. In the use cases reported here, the current-state model corresponded to a disease-state immune-cell model and the target-state model corresponded to a matched healthy reference model. Users also provide a natural-language biological question through the web interface.

Once MechAInistic receives the input files and the question, the user may submit the analysis request. The interface also exposes configurable settings for the LLMs and metabolic model objective functions, allowing users to adjust key analysis parameters before execution. By default, MechAInistic populates the LLM backend setting with the National Research Platform's (NRP) OpenAI-compatible endpoint, which is free for academic researchers to use upon signing up with NRP.[35] Alternatively, users can provide a different URL endpoint (e.g., OpenAI, Anthropic) and API key, provided the endpoint is publicly accessible; additional information on setting up MechAInistic with local LLMs is available in the Supplementary Information section titled “LLM Configuration.” Figure 1 shows the adjustable configuration settings and the web interface after the modeling files have been uploaded and the system is ready to process the request.

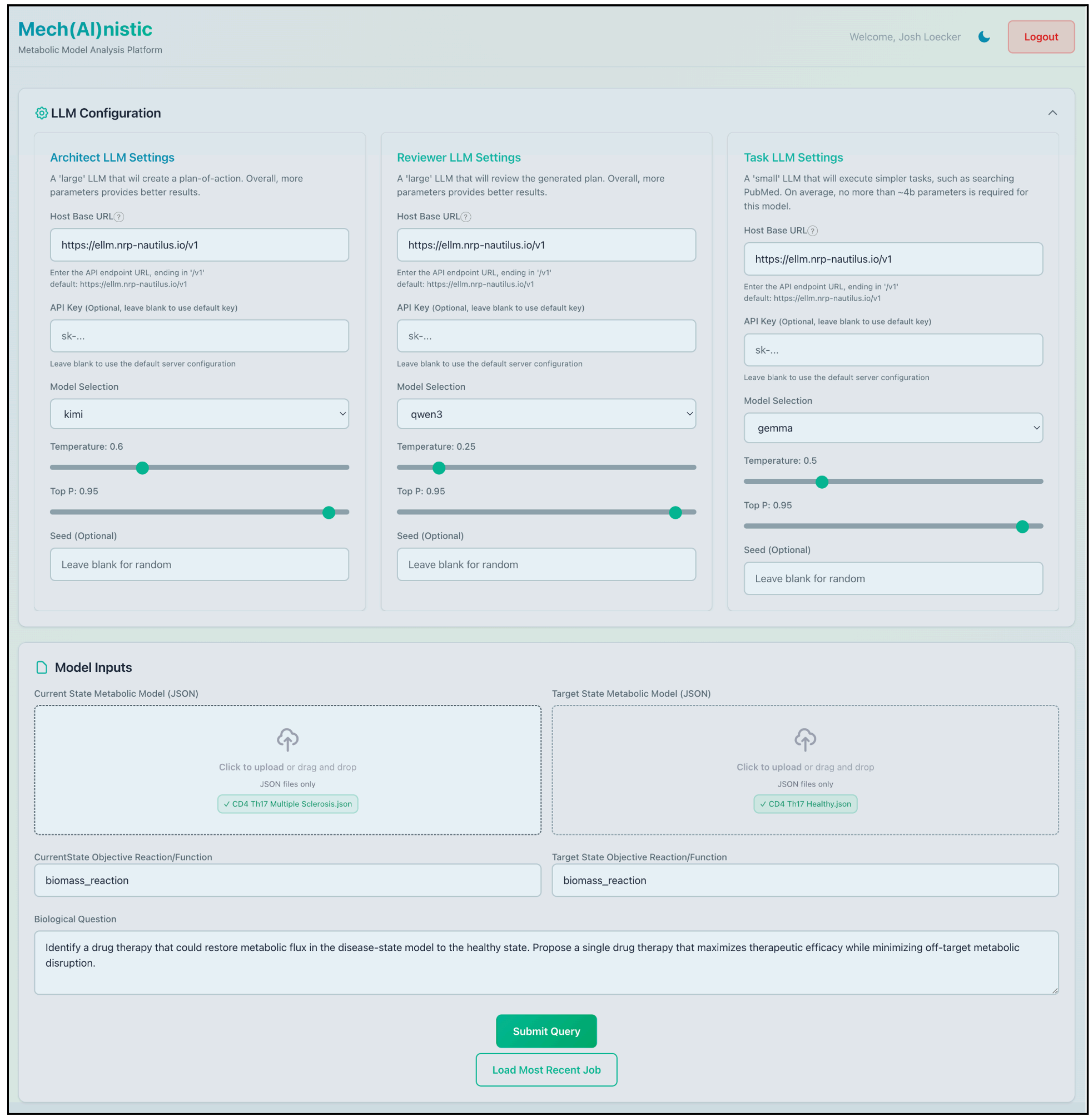

**Figure 1**: The adjustable LLM settings and model upload section of MechAInistic. The current-state (disease state) and target-state (healthy state) models have been uploaded, and the system is ready to process requests.

After submission, MechAInistic creates a session-specific analysis state containing the uploaded models, user query, selected objective functions, LLM configurations, available tools, accumulated tool calls, summarized tool output, and final report state. Reports are rendered in the browser and can be downloaded in Markdown format.

Architecturally, MechAInistic is implemented as a modular, containerized system, where the backend, frontend, a PostgreSQL database, and LLM host can each run as independent services on separate servers or nodes. This architecture allows the platform to be deployed across one or more machines, provided that each component can communicate with its required dependency. In the current implementation, the main backend communicates with several external services during execution: the LLM host, external databases (PubMed, DGIdb, MyGeneInfo, etc.), and a PostgreSQL database. This separation supports flexible deployment on local or distributed hardware while preserving the same high-level workflow exposed through the web interface. Figure 3 illustrates the system's deployment architecture.

### *Agent roles and workflow control*

MechAInistic uses three role-specialized agents: Architect, Reviewer, and Task. The Architect is responsible for generating an initial plan, synthesizing tool calls, and constructing a final report, while the Reviewer will provide feedback and suggestions to improve the analysis and interpretation during each step. Additionally, the Reviewer will score the analysis and serve as a stopping point to prevent the Architect from finalizing it without sufficient quantitative support from literature or the metabolic models. The Task agent supports two auxiliary functions: translating biological hypotheses into PubMed-oriented search queries and summarizing raw tool outputs into concise plain-text summaries that preserve quantitative values, identifiers, caveats, and errors for downstream reasoning. In this way, MechAInistic provides coordinated multi-agent execution while helping preserve the consistency and fidelity of the analytical process and its associated outputs.

After users submit their query, , MechAInistic initializes a session-specific analysis state containing the uploaded data and available tools and creates placeholders for completed tool calls, summarized results, and Reviewer feedback (Figure 2A). The workflow follows a multi-stage, multi-agent design with "outer" and "inner" control systems. The outer loop has a non-user-configurable maximum of 5 retries to prevent infinite execution of the Agents; the inner loop has a maximum of 10 iterations, each executing 1 to 6 tool calls, to prevent infinite loops. The retry logic, tool calling state management, and agent outputs are defined in Supplementary Tables S4-S7. Following this, the workflow then proceeds through three main phases: Plan Development, Iterative Synthesis and Execution, and Report Generation.

First, during **Plan Development** (Figure 2B), the Architect model receives the user's biological query, a list of available analysis tools, and instructions to generate an experimental, in-silico workflow that, when executed, answers the user's question. Once the plan is constructed, the Reviewer evaluates the plan and returns feedback in the following format: (a) a score on a scale from 1 to 10, (b) feedback describing the reason for the provided score, and (c) a list of suggestions to improve the plan. The Reviewer assessment is based on five components, each worth two points: Completeness (does the plan address all aspects of the user's question?), Tool Appropriateness (are the selected tools suitable for the task?), Logical Sequence (are the steps performed in the correct order?), Feasibility (can the proposed plan be completed?), and Error Handling (are failure cases addressed?). The reviewer assigns each category a score of 0 (not answered), 1 (partially answered), or 2 (fully answered). The scores are summed, and if the result is 7 or lower, the plan is considered a failure and must be regenerated. In this case, the Architect will utilize the Reviewer provided suggestions when generating a new plan. If the plan passes, the workflow continues to the Iterative Synthesis and Execution phase.

During the **Iterative Synthesis and Execution** phase (Figure 2C), the Architect is provided with their approved plan and the Reviewer's feedback (regardless of whether the original plan passed or failed), and begins synthesizing tool calls as defined by the analysis. These tools are returned to MechAInistic's System Orchestrator, which parses the Architect's response and sequentially executes the tool calls. After execution, the Task model is used to summarize results; token usage decreases by approximately 50% when structured JSON results are summarized into plain text. After summarization, the Reviewer acts as an intermediary to determine if additional tool calls are required to answer the user's query. The Reviewer provides the following feedback: (a) a score on a scale from 1 to 10, (b) feedback describing what has been accomplished thus far, and (c) a list of requirements for improvement, including what is missing, potential next steps, or refinements for the analysis (Supplementary Table S6). Because the Iterative Synthesis phase can call multiple tools during its cycle, the scoring is more lenient to provide the Architect additional chances for correction. The tool calling results are assigned a score of 1 to 3 if there was a malfunction (e.g., software crash, wrong format, incomplete execution); 4 or 5 if the tool was executed correctly but the results are unusable (e.g., nonsensical output format or results do not correlate to the goal); 6 or 7 if the tool was executed properly and the results appear to be related to the goal but require additional investigation; or an 8 to 10 if the tool was executed correctly, the results are biologically plausible, and ready to be used in the final report generation. During this phase, there is a maximum limit of 10 loops to prevent infinite tool calling. If the Architect signals no further steps are needed, or if the Reviewer determines that the analysis is complete, the Iterative Synthesis loop terminates. As described previously, during this phase, the initial plan will be expanded if it proves insufficient to answer the user's query.

Importantly, low-scoring Iterative Synthesis phases will not increment the outer loop's retry count, as both the Architect and Reviewer must independently indicate that the result is finalized. It is only at this point that the entire tool calling results are provided to the Reviewer, where a second, broader determination is made. At this point, if the Reviewer determines that the results are not complete, the retry counter is incremented.

In the **Result Evaluation** phase (Figure 2D), the Reviewer is provided the all summarized tool call results and decides if enough information has been obtained to answer the user's query. If it is determined that yes, all components are answered and the initial plan is complete, the workflow continues to the Report Generation phase. Otherwise, the outer loop retry count is incremented by one and the Iterative Synthesis phase is re-entered.

Finally, in the **Report Generation** phase (Figure 2E), the Architect is given the entire workflow state, including the user's query, the original plan, all summarized tool call results, and synthesizes a final report of all findings, including their mechanistic interpretation. The report is displayed in the browser and can be downloaded in Markdown format.

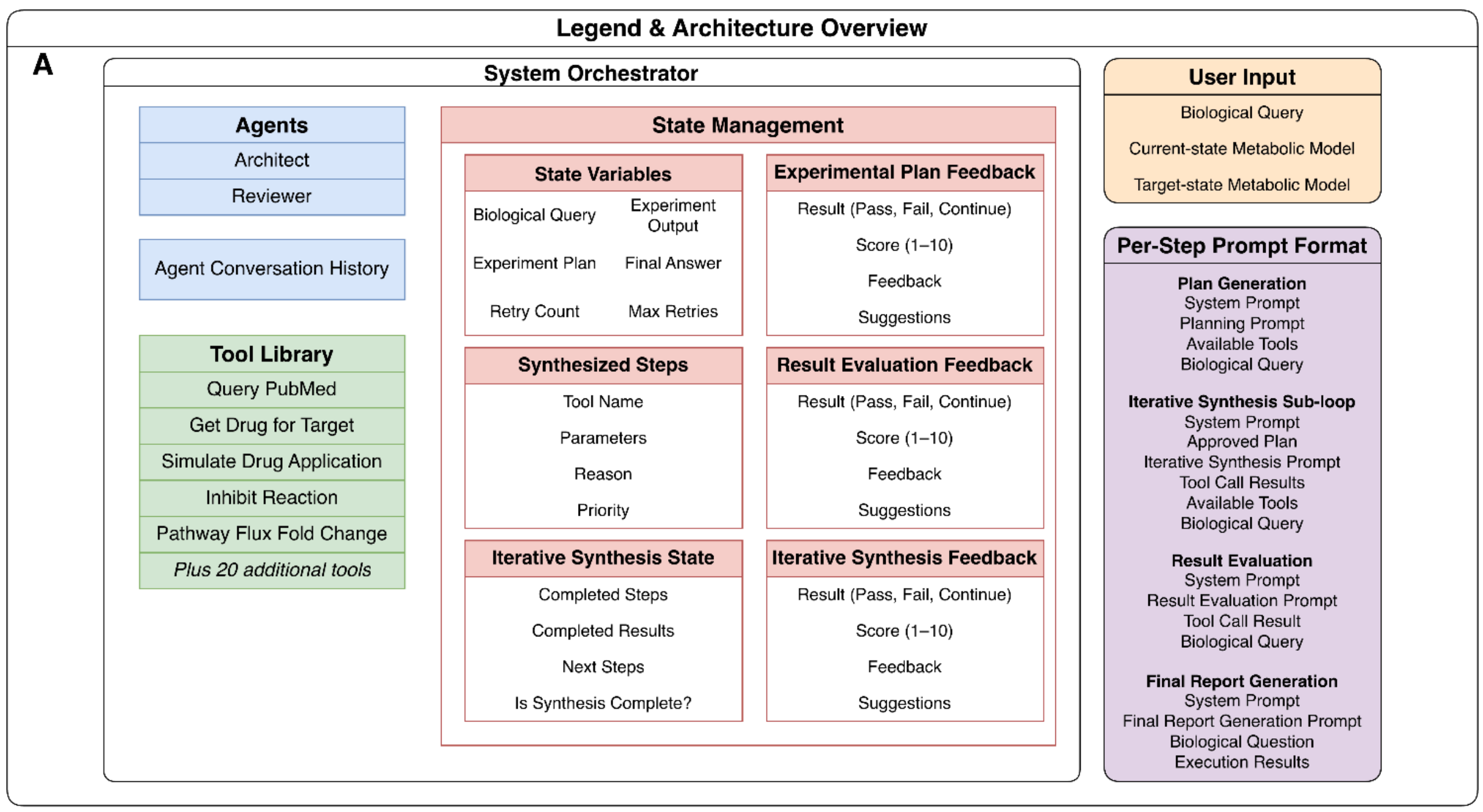
Legend & Architecture Overview
A
System Orchestrator
Agents
Architect
Reviewer
Agent Conversation History
Tool Library
Query PubMed
Get Drug for Target
Simulate Drug Application
Inhibit Reaction
Pathway Flux Fold Change
Plus 20 additional tools
State Management
State Variables
Biological Query
Experiment Output
Experiment Plan
Final Answer
Retry Count
Max Retries
Experimental Plan Feedback
Result (Pass, Fail, Continue)
Score (1–10)
Feedback
Suggestions
Synthesized Steps
Tool Name
Parameters
Reason
Priority
Result Evaluation Feedback
Result (Pass, Fail, Continue)
Score (1–10)
Feedback
Suggestions
Iterative Synthesis State
Completed Steps
Completed Results
Next Steps
Is Synthesis Complete?
Iterative Synthesis Feedback
Result (Pass, Fail, Continue)
Score (1–10)
Feedback
Suggestions
User Input
Biological Query
Current-state Metabolic Model
Target-state Metabolic Model
Per-Step Prompt Format
Plan Generation
System Prompt
Planning Prompt
Available Tools
Biological Query
Iterative Synthesis Sub-loop
System Prompt
Approved Plan
Iterative Synthesis Prompt
Tool Call Results
Available Tools
Biological Query
Result Evaluation
System Prompt
Result Evaluation Prompt
Tool Call Result
Biological Query
Final Report Generation
System Prompt
Final Report Generation Prompt
Biological Question
Execution Results

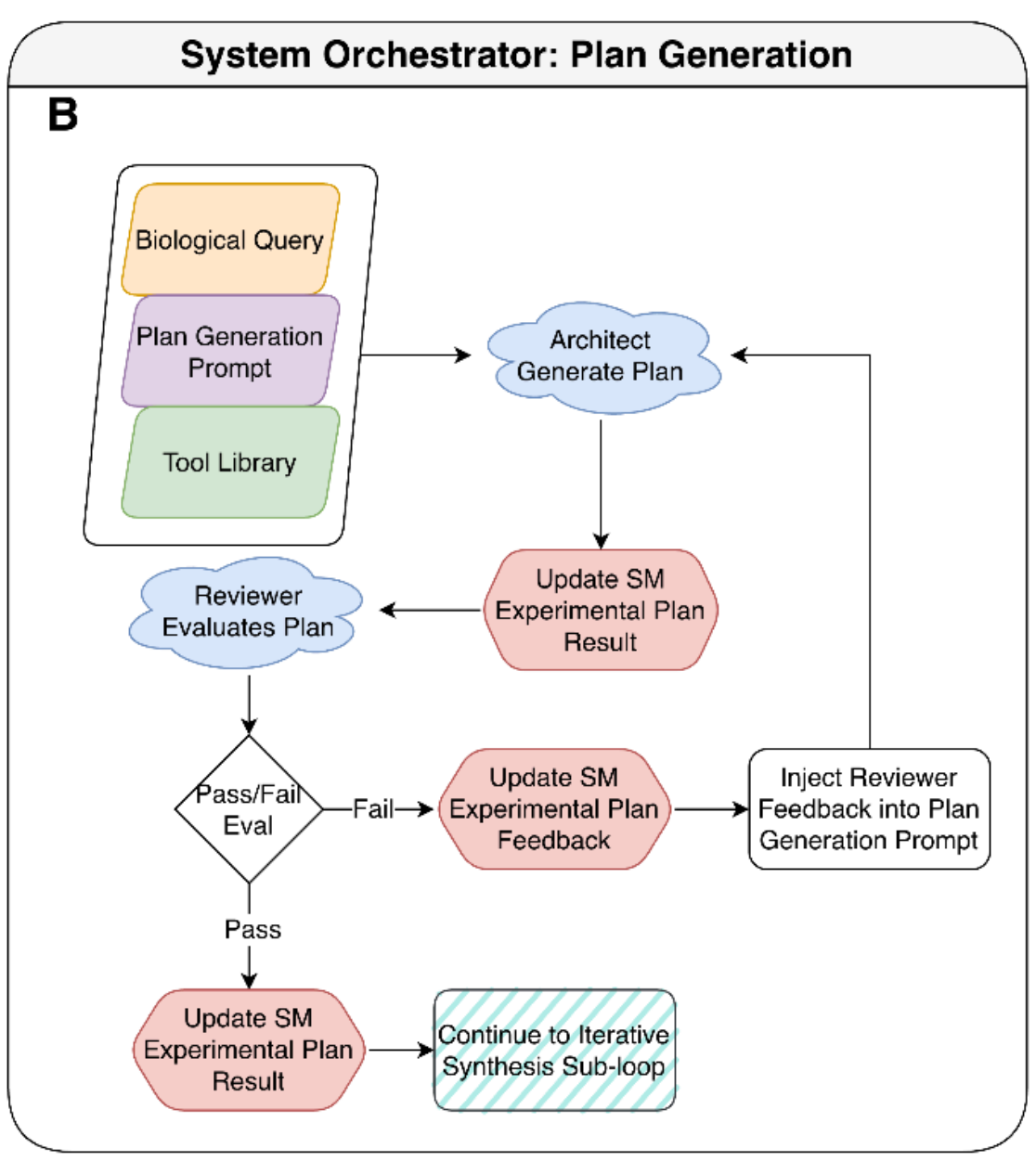
System Orchestrator: Plan Generation
B
Biological Query
Plan Generation Prompt
Tool Library
Architect Generate Plan
Update SM Experimental Plan Result
Reviewer Evaluates Plan
Pass/Fail Eval
Fail
Update SM Experimental Plan Feedback
Inject Reviewer Feedback into Plan Generation Prompt
Pass
Update SM Experimental Plan Result
Continue to Iterative Synthesis Sub-loop

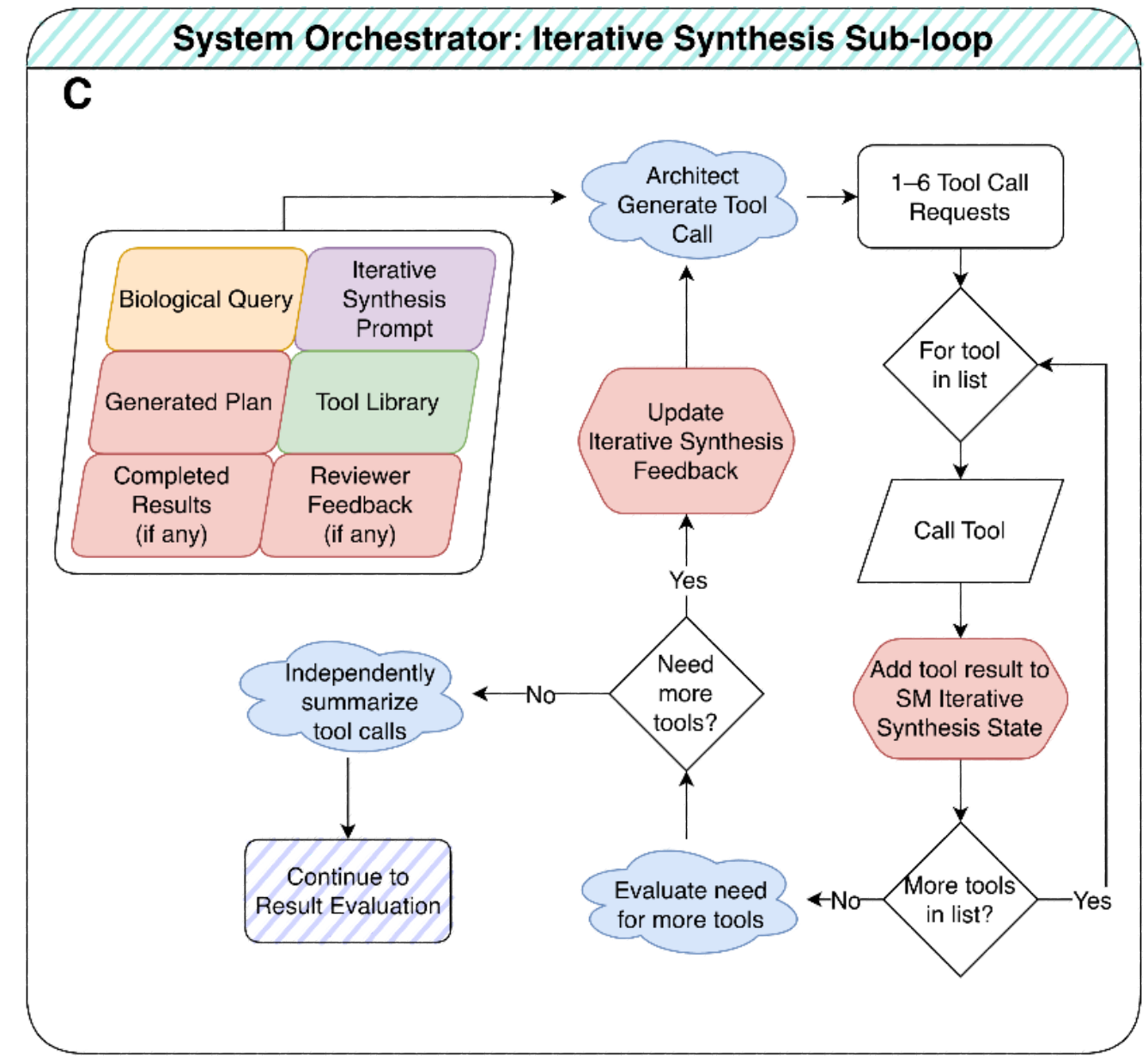
System Orchestrator: Iterative Synthesis Sub-loop
C
Biological Query
Iterative Synthesis Prompt
Generated Plan
Tool Library
Completed Results (if any)
Reviewer Feedback (if any)
Architect Generate Tool Call
1–6 Tool Call Requests
For tool in list
Call Tool
Add tool result to SM Iterative Synthesis State
More tools in list?
Yes
No
Evaluate need for more tools
Need more tools?
Yes
Update Iterative Synthesis Feedback
No
Independently summarize tool calls
Continue to Result Evaluation

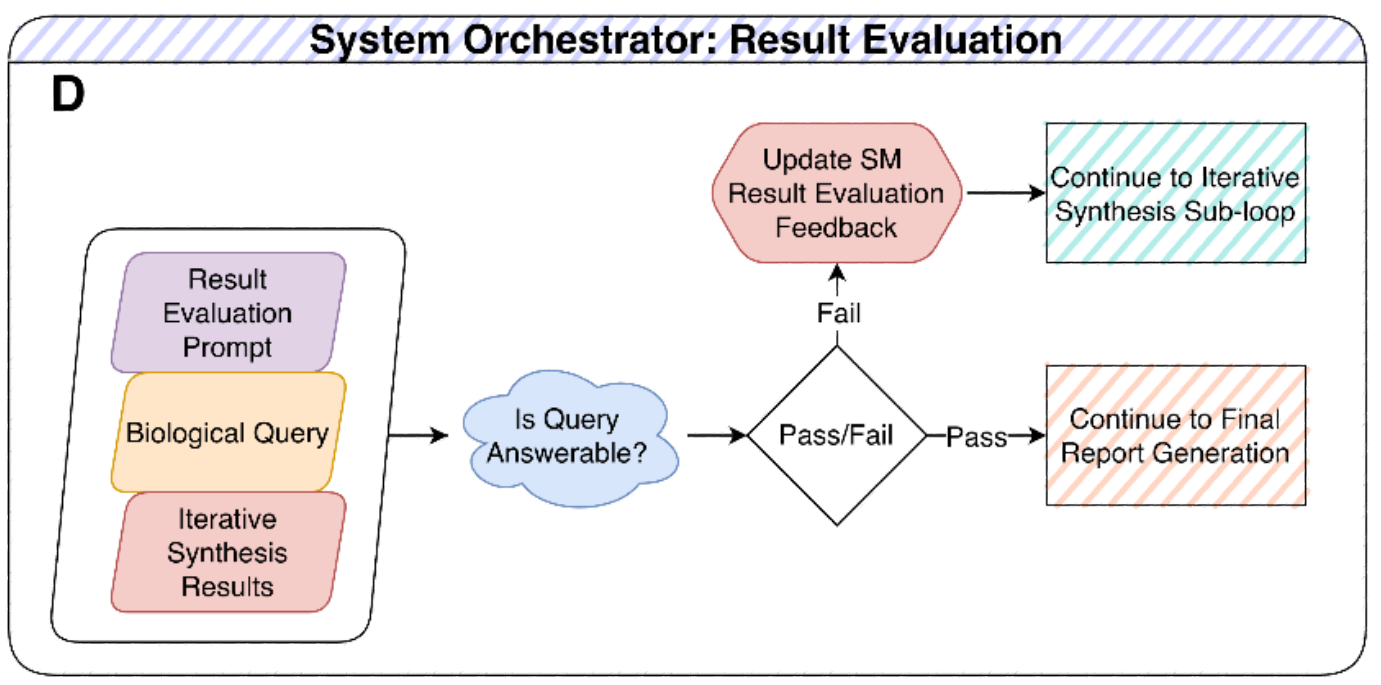
System Orchestrator: Result Evaluation
D
Result Evaluation Prompt
Biological Query
Iterative Synthesis Results
Is Query Answerable?
Pass/Fail
Fail
Update SM Result Evaluation Feedback
Continue to Iterative Synthesis Sub-loop
Pass
Continue to Final Report Generation

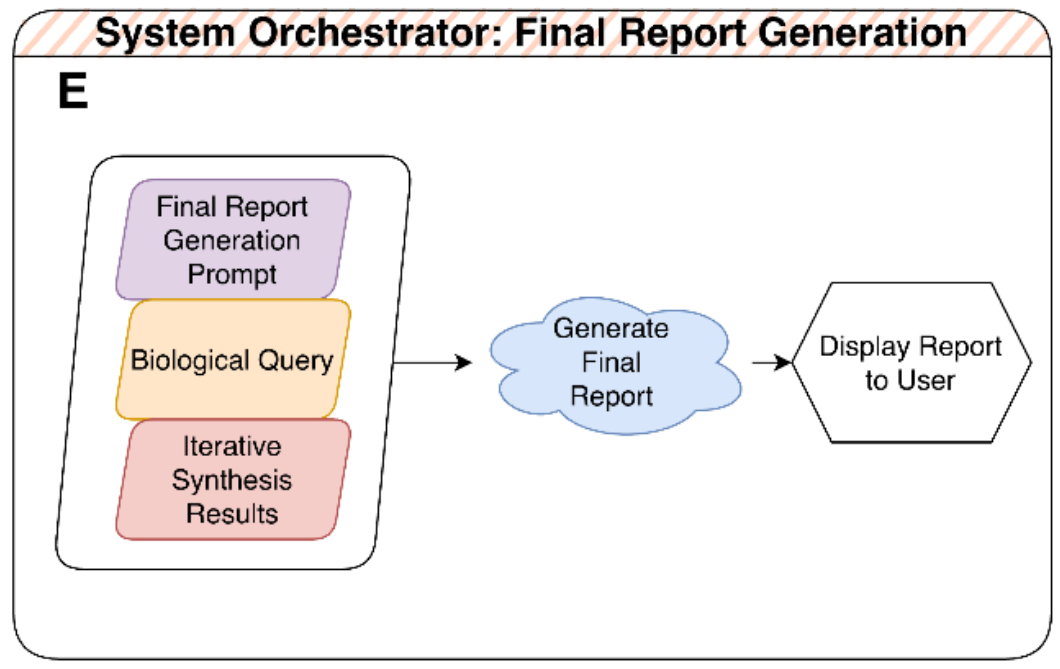
System Orchestrator: Final Report Generation
E
Final Report Generation Prompt
Biological Query
Iterative Synthesis Results
Generate Final Report
Display Report to User

**Figure 2**: Overview of the MechAInistic system architecture. **Figure 2A)** The legend and diagram overview. Colors indicated by this panel are replicated throughout the figure. The primary agents include the Architect and Reviewer, each with their own isolated conversation history; the Task agent is not depicted here as its role is limited to processing the input and output of specific tool calls. **Figure 2B)** The Plan Generation information flow. The Architect agent receives the user's Biological Query, the Plan Generation Prompt, and a list of tools plus their descriptions. The Reviewer evaluates the plan; if the plan succeeds, State Management is updated; otherwise, the Architect re-attempts plan generation. **Figure 2C)** The Iterative Synthesis Sub-loop. The architect receives the user's biological query, the Iterative Synthesis Prompt, the generated plan, all available tools, completed results (if available), and Reviewer agent feedback (if available). **Figure 2D)** The Result Evaluation process. The Reviewer agent receives the Result Evaluation Prompt, the user's biological query, and all results from the Iterative Synthesis Sub-loop. If the Reviewer agent indicates that the results are not sufficient to answer the user's query, the Iterative Synthesis Sub-loop is re-entered (Panel C); otherwise, a final report will be generated. **Figure 2E)** Final Report Generation. The Architect receives the Final Generation Report, the user's Biological Query, and all Iterative Synthesis Results. The Architect synthesizes a structured report in Markdown format, which is displayed to the user.

### *LLM configuration and prompt management*

The platform employs three large language models. Each model is configured at runtime using a common parameter set: model name, endpoint host, API key, temperature, top-p sampling value, and an optional random seed. This allows MechAInistic to make use of any LLM provider across any combination of hosts. When the user submits their biological question, these parameters are sent to the backend and passed to an OpenAI-compatible client interface, enabling the system to interact with externally hosted models via a standardized API. As described previously, we are using NRP-hosted language models as the default LLM provider. We chose separate models to serve as the Architect and Reviewer to incorporate model diversity into our evaluation pipeline, leverage their distinct training methodologies, and exploit their different token output probabilities. The Task model was chosen for its fast inference speed. The temperature for each agent was set to 0.5 to balance output consistency with some degree of variability during tool calls, reviewing, and report generation.

The Architect, Reviewer, and Task agents are given deliberately asymmetric, role-specific instructions. The Architect is framed as a strategy-oriented analytical agent, emphasizing methodological transparency, evidence-based planning, risk anticipation, tool orchestration, and failure-mode mitigation. The Reviewer is framed as a rigorous, yet constructive, evaluator, emphasizing the identification of methodological weaknesses, balancing strengths and limitations, and providing actionable feedback. The Task agent's function is two-fold; first, it finds academic publications from PubMed and translates complex biological hypotheses into optimized search queries that balance precision with recall, ensuring relevance and adequate result coverage. Second, it summarizes tool call results from a JSON format into plain text. This prompt asymmetry operationalizes the division of labor between analytical generation, critical validation, and retrieval-augmented generation.

The execution layer exposes a curated set of metabolic analysis tools through a central registry in the agent pipeline. The tools are registered in the system as callable functions and are made available to the Architect agent during both the planning and iterative synthesis phases, so the Architect is explicitly informed of the available computational actions and their intended use, rather than relying solely on prior knowledge. For each available tool, MechAInistic dynamically extracts the tool name, a 1-sentence description, and parameters required to call the tool. Examples of these parameters include a reaction, pathway, gene name, or a drug to apply to the metabolic model. The toolset supports several categories of analysis, including metabolic model simulation, reaction- and pathway-level flux analysis, graph-based interrogation of metabolic networks, intervention and perturbation analysis, metabolite lookup, gene knockout simulation, drug-target discovery, and literature retrieval.

To manage prompts for each phase, templates are organized by functional stage and agent role, then instantiated through placeholder substitution and summarized execution results at runtime. This design separates control logic from prompt text and supports role-specific prompting for planning, execution, evaluation, and final answer generation. Within each phase, a System Prompt primes the respective Agent for its task. These prompts are intended to serve as a computational methods template rather than a simple instruction; they require explicit treatment of assumptions, boundary conditions, workflow modularity, validation checkpoints, failure modes, sensitivity analysis, and evidence integration. Supplementary Table S4 summarizes the prompts for each phase and their expected outputs. The full prompt templates are provided in Supplementary Information Note 10 for transparency and reproducibility.

### *Tool registry and executable analysis layer*

The execution layer exposes a curated registry of metabolic-analysis tools to the Architect agent. Each tool is registered with a name, short description, required parameters, optional parameters, and expected input types. During Plan Development and Iterative Synthesis, MechAInistic provides the Architect with the available tool descriptions so that proposed workflows are constrained to executable operations rather than unconstrained language-model reasoning. When the Architect proposes a tool call, it must return the exact tool name and complete parameter assignments required for execution.

The backend parses the Architect's proposed tool calls, validates required parameters, injects the session-specific model/cache key, and executes the corresponding function. This session key identifies the uploaded current-state and target-state models and is handled by the backend rather than exposed to the LLM. Raw tool outputs are stored in the session state, then summarized by the Task agent before being returned to the Architect and Reviewer for subsequent reasoning. This design preserves the complete computational record while reducing the token burden of passing large structured outputs between agents.

The tool registry supports several categories of analysis: metabolic-model simulation, flux-distance calculation, reaction-level flux queries, pathway-level comparison, graph/network analysis, robustness analysis, gene and reaction perturbation, drug-effect simulation, dose-response analysis, drug-target lookup, metabolite lookup, and literature retrieval. These tools allow MechAInistic to move from a natural-language query to model-derived quantities, candidate reaction prioritization, perturbation testing, drug mapping, and literature-grounded interpretation.

For the use cases reported here, the executed workflows used tools for baseline flux comparison, pathway-level dysregulation, reaction prioritization, robustness evaluation, simulated inhibition, dose-response analysis, drug-target lookup, and PubMed retrieval. Full tool descriptions, parameter definitions, and representative tool categories are provided in Supplementary Tables S3 and S4.

### *Constraint-based modeling backend*

The primary CBM software stack is built around CobraPy, the HiGHS[36] solver, and additional supporting libraries such as NetworkX for graph-based analysis and Pandas for flux and result processing (Table 2, Supplementary Table S8).

| CBM Component | System Interpretation |
| --- | --- |
| Core Modeling Library | COBRApy |
| LP Solver | HiGHS |

| CBM Component | System Interpretation |
| --- | --- |
| Model representation | JSON format |
| Primary Simulation Type | Flux Balance Analysis |
| Additional analyses | Reaction/pathway flux comparison, robustness analysis, gene/reaction knockout, graph analysis, drug-effect simulation, dose-response, flux-transition optimization |
| Objective specification | User-supplied at runtime; can be changed later with a dedicated tool by the Architect Agent |
| Constraint handling | Inherited from the model; can be adjusted later through tool-specific modifications by the Architect Agent |

**Table 2**: Constraint-based modeling software and analysis configuration

The dominant analysis paradigm in the current system is flux balance analysis, which serves as the main simulation routine for each metabolic model. In addition to standard steady-state optimization, MechAInistic includes reaction-level and pathway-level flux comparisons, robustness (Phenotype Phase Plane) analysis using COBRApy's production envelope function[37], and perturbation analysis via single-gene, single-reaction, double-gene, and double-reaction knockouts. The repository also provides graph-based analyses of the metabolic models and intervention-oriented routines such as simulated drug effects, dose-response analysis, and optimization of reaction modifications to reduce flux-distance between disease and target states.

### *Metabolic model reconstruction and use-case setup*

We evaluated MechAInistic using two paired immune-cell metabolic-model use cases. (1) Naive B models in a Healthy (as described previously)[38] and Rheumatoid Arthritis (RA) disease state, and (2) CD4+ Th17 models in a Healthy (as described previously)[39] and Multiple Sclerosis (MS) disease state. To reconstruct the RA and MS disease models, bulk RNA-seq data were obtained from the NCBI Gene Expression Omnibus.[40] The Naive B RA datasets were very limited, and only a single study containing four samples of bulk RNA-seq RA data was identified.[41] For the MS reconstruction, 19 samples from a single study were identified. All samples were preprocessed using AutoRNAseq[42], which aligned the data to the Homo Sapiens reference genome (Ensembl Release version 115[43]) using the STAR[44] aligner and quantified reads using Salmon[45]. Following preprocessing, exchange reactions for both the RA and MS conditions were constrained using appropriate disease-specific constraints, as well as lower and upper bounds. The model reconstruction was performed using COMO[38], a Jupyter Notebook-based tool designed to automate and ensure reproducibility in constraint-based metabolic model reconstruction, with Recon3D[12] as the reference model. The Supplementary Material section titled “Constraint-Based Model Reconstruction” contains details for all accession values, exchange reactions, and reaction bounds for model reconstruction.

The same natural-language prompt was used for both use cases: “Identify a drug therapy that could restore metabolic flux in the disease-state model to the healthy state. Propose a single drug therapy that maximizes therapeutic efficacy while minimizing off-target metabolic disruption.” The disease and cell-type labels were not included in the prompt and were available to MechAInistic only through the uploaded model context.

### *Comparator LLM evaluation*

To compare MechAInistic with general-purpose LLM workflows, we submitted the same paired metabolic-model files and the same drug-therapy prompt to Anthropic's Claude, OpenAI's ChatGPT, Microsoft's Copilot, and Google's Gemini. Each system was asked to identify a single therapy that would restore disease-state metabolic flux toward the healthy reference while minimizing off-target metabolic disruption. Comparator outputs were saved as standalone reports and evaluated using the nine-axis rubric described below.

Because general-purpose LLM systems differ in file ingestion capabilities, code execution environments, tool access, context length limits, and access to external literature, we evaluated the returned artifacts rather than assuming equivalent computational access. The assessment therefore focused on whether each output provided evidence that the uploaded models were used, whether quantitative model-derived values were reported, whether a specific therapy was recommended, whether literature support was traceable, and whether the reasoning path was reproducible and auditable.

The identity of the recommended drug was recorded as a descriptive outcome, but was not treated as the primary measure of correctness. Because the correct therapeutic answer cannot be established from computational analysis alone, the primary comparison focused on model grounding, quantitative evidence, methodological traceability, task completion, and explicit recognition of limitations. Full comparator outputs and per-system scoring justifications are provided in Supplementary Information Note 8.

### *Nine-axis evaluation rubric*

MechAInistic and comparator outputs were assessed using a nine-axis rubric designed to distinguish plausible biological narrative from model-grounded computational reasoning. The rubric evaluated whether each system produced an answer that was not only biologically plausible but also traceable to the supplied metabolic models, supported by quantitative evidence, and aligned with the user's requested task.

The nine evaluation axes were:

1. **Model fidelity:** evidence that the uploaded metabolic models were loaded, parsed, and used in the analysis.
2. **Methodological reproducibility:** inclusion of named tools, solvers, scoring rules, executable analysis steps, or other information needed to reproduce the workflow.
3. **Literature evidence:** use of traceable peer-reviewed sources, including PMIDs, DOIs, or clearly identifiable references where applicable.
4. **Computational evidence:** inclusion of flux values, perturbation outputs, distance metrics, dose-response results, or other quantitative model-derived outputs.
5. **Direct relevance:** adherence to the single-drug flux-restoration task rather than drifting into general disease biology or unrelated therapeutic framing.
6. **Question completion:** whether the system returned a specific single therapy in response to the prompt.
7. **Structured review or agentic control:** evidence of a review, checkpointing, or multi-agent control process before final synthesis.

8. **Execution feasibility and rerunnability:** completion without file-ingestion failure, truncation, unavailable context, or undocumented execution constraints.
9. **Human-review traceability:** explicit indication of manual review, sign-off, correction, or identifiable points where human review could be applied.

Each axis was scored qualitatively as full, partial, or failed based on pre-specified criteria. Drug identity was recorded descriptively but was not used as the main correctness criterion, because the computational analyses generate hypotheses rather than validated therapeutic answers. The primary evaluation focused on whether each recommendation was supported by model-grounded computation, quantitative evidence, methodological traceability, and explicit limitations.

### *Manual validation of MechAInistic outputs*

Manual validation was performed after MechAInistic generated the final reports for each use case. The goal of this validation was not to prove therapeutic correctness, but to determine whether the model-derived results, literature-supported claims, and final interpretations were traceable and internally consistent.

Validation focused on five categories: (1) whether reported quantitative values matched the underlying tool-call outputs; (2) whether reaction identifiers, pathway names, drug names, and model states were transcribed correctly; (3) whether literature cited by MechAInistic could be traced to real publications; (4) whether the cited literature supported the specific mechanistic claim made in the report; and (5) whether the final therapeutic interpretation followed from the model-derived and literature-supported evidence.

Validation outcomes were classified as supported, partially supported, unsupported, or requiring expert interpretation. Supported findings were retained in the manuscript. Partially supported findings were retained only with explicit caveats. Unsupported claims, citation mismatches, annotation-related errors, or cases requiring expert reinterpretation were reported as failure modes rather than removed from the analysis. This procedure was used to distinguish model-grounded therapeutic hypotheses from validated therapeutic recommendations.

### *External services and reproducibility*

The analytical pipeline calls external scientific services during execution, including BiGG[46], DGIdb[47], MyGeneInfo[48], and PubMed[49]. These services support reaction and metabolite annotation, drug-gene interaction lookup, gene identifier resolution, and literature retrieval. MechAInistic also maintains local infrastructure for user/session data, drug-target caching, and literature-embedding search. External-service dependencies, rate limits, caching behavior, and software dependencies are described in the Supplementary Information.

## DISCUSSION

We developed MechAInistic as a supervised, tool-grounded agentic system for reasoning over paired constraint-based metabolic models. Across two immune-cell use cases, the system converted natural-language prompts into executable workflows, generated model-derived quantitative evidence, and produced mechanistic therapeutic hypotheses with explicit limitations. These results support a role for large language models not as unconstrained biological answer engines, but as orchestration layers for executable mechanistic modeling workflows.

A central finding is that workflow structure matters.[50] General-purpose LLM systems can generate biologically plausible explanations and, in some cases, useful therapeutic suggestions. However, plausibility alone is insufficient for model-driven biological reasoning. The key question is whether the final answer can be traced to uploaded models, executable tool calls, quantitative outputs, and literature-supported interpretation. MechAInistic's Architect-Reviewer-Task architecture addresses this need by separating planning, execution, intermediate review, literature retrieval, and final synthesis.

The Architect-Reviewer pattern was motivated by a broader challenge in LLM-enabled biomedical reasoning: language models can generate fluent but unsupported biological claims unless constrained by external verification. Prior work has shown that biomedical associations generated by LLMs can be improved when terminology is validated against specialized ontologies or databases, factual claims are checked against biomedical literature, and outputs are reviewed by independent models or verification workflows.[50–56] MechAInistic applies these principles to metabolic modeling by requiring the Architect's proposed analytical path and final interpretation to pass through executable CBM tools, Reviewer-mediated critique, and literature-grounded synthesis.

The two use cases illustrate MechAInistic's role in therapeutic hypothesis generation. In rheumatoid arthritis Naive B cells, the system identified mitochondrial metabolic rewiring and nominated Devimistat/CPI-613 as an investigational OGDH-centered hypothesis. In multiple sclerosis CD4+ Th17 cells, it identified ICDHy/IDH as a candidate metabolic control point and proposed IDH inhibition as a therapeutic hypothesis. These outputs should not be interpreted as validated therapeutic recommendations; rather, they are model-grounded hypotheses for expert review and experimental validation.

Several limitations remain. First, MechAInistic depends on the quality, completeness, and biological validity of uploaded metabolic models. Missing reactions, genes, metabolites, exchange constraints, or annotations are not automatically repaired, so downstream analyses remain bounded by the input models. In this study, the RA and MS disease-state models were reconstructed from limited bulk RNA-seq datasets, and the models themselves require separate documentation and validation. Second, although tool calls provide a grounding layer, LLM behavior can vary across runs, model versions, providers, and inference settings. Reproducible use therefore requires reporting uploaded models, prompt text, objective functions, LLM configuration, solver, tool versions, and execution date.

Third, literature retrieval and summarization require verification. Manual validation identified at least one citation-level mismatch despite retrieval of biologically relevant literature, reinforcing the need to verify citations, drug-target claims, and mechanistic interpretations before downstream use. Such errors are consistent with broader limitations of LLM-mediated summarization and factuality across biomedical and scientific contexts.[57–61] Fourth, the current interface supports self-contained analyses rather than open-ended multi-turn scientific conversations, and complex workflows require substantial token usage; token counts are provided in Supplementary Table S15. Finally, MechAInistic depends on external biological databases and literature services, including BiGG[46], DGIdb[47], MyGeneInfo[48], and PubMed[49], so incomplete or unavailable resources may degrade drug-target mapping or literature-grounded interpretation.

Together, these findings establish MechAInistic as a step toward natural-language access to executable mechanistic biology. The platform does not replace expert computational modeling or experimental validation. Instead, it provides an auditable interface between biological questions, CBMs, computational tools, and literature-supported interpretation.

## FUTURE DIRECTIONS

MechAInistic establishes a foundation for natural-language-driven reasoning over constraint-based metabolic models, and several avenues remain open for further development. Continued work will focus on broadening the range of supported analyses, evaluating the system across additional cell types and disease contexts, and incorporating advances in language model architectures as they become available. Because the language model field is continuously advancing, we envision future updates that auto-route the user's query to the most favorable cost-to-performance language model based on what is available at the time. We also anticipate ongoing refinement of the multi-agent framework informed by community feedback and user experience. Additional extensions to the platform are under active development and will be described in subsequent work.

## ACKNOWLEDGMENTS

This work was supported by the National Institutes of Health under Grant No. #R35GM119770, and the University of Nebraska-Lincoln Grand Challenges Catalyst Award to Tomáš Helikar.

We are thankful to Skylar Loecker and Robert Moore II for their valuable discussions during the methodology and investigation phases of this project.

## AUTHOR CONTRIBUTIONS

Conceptualization, T.H.; methodology, J.L., N.P, W.B., P.P., A.A.H., T.H.; Investigation, J.L., P.P., B.L.P., A.A.H., T.H.; writing—original draft, J.L., A.A.H., T.H.; writing—review & editing, J.L., P.P., B.L. P., A.A.H., T.H.; funding acquisition, T.H.; resources, T.H.; supervision, P.P., A.A.H., T.H.;

## DECLARATION OF INTERESTS

T.H is a founder and shareholder of Discovery Collective, Inc., and ImmuNovus, Inc.

## DECLARATION OF GENERATIVE AI AND AI-ASSISTED TECHNOLOGIES

During the preparation of this work, the authors used MechAInistic (an LLM-guided multi-agent system built on large language models including Qwen, Kimi, and Gemma) to execute all constraint-based metabolic analyses, generate analytical plans, interpret flux-based results, and produce structured reports for the use cases described in this manuscript. Claude (Anthropic) was used to reformat and restructure these outputs into manuscript-ready sections. All AI-generated content was reviewed, edited, and validated by the authors, who take full responsibility for the scientific content.